\documentclass[sigconf]{acmart}
\usepackage{xcolor}
\usepackage{dsfont}
\usepackage{float}
\usepackage[colorinlistoftodos,prependcaption,textsize=tiny,linecolor=red,backgroundcolor=red!25,bordercolor=red]{todonotes}
\def\BibTeX{{\rm B\kern-.05em{\sc i\kern-.025em b}\kern-.08emT\kern-.1667em\lower.7ex\hbox{E}\kern-.125emX}}
    
\copyrightyear{2019}
\acmYear{2019}
\setcopyright{acmlicensed}
\acmConference[SIGIR '19]{SIGIR '19: The 42nd International
ACM SIGIR Conference on Research \& Development in Information Retrieval}{July 21--25, 2019}{Paris, France}
\acmBooktitle{SIGIR '19: The 42nd International
ACM SIGIR Conference on Research \& Development in Information Retrieval, July 21--25, 2019, Paris, France}
\begin{document}

%
% The "title" command has an optional parameter, allowing the author to define a "short title" to be used in page headers.
\title{Global Aggregations of Local Explanations for Black Box models}

%
% The "author" command and its associated commands are used to define the authors and their affiliations.
% Of note is the shared affiliation of the first two authors, and the "authornote" and "authornotemark" commands
% used to denote shared contribution to the research.
\author{Ilse van der Linden}
\affiliation{%
  \institution{University of Amsterdam}
  \streetaddress{Science Park 904}
  \city{Amsterdam}
  \state{The Netherlands}
}
\email{iwc.vanderlinden@gmail.com}
%\email{}
% ORCID???

\author{Hinda Haned}
\affiliation{%
  \institution{University of Amsterdam, \\ Ahold Delhaize}
  \streetaddress{Science Park 904}
  \city{Amsterdam}
  \country{The Netherlands}
}
\email{h.haned@uva.nl}

\author{Evangelos Kanoulas}
\affiliation{%
  \institution{University of Amsterdam}
  \streetaddress{Science Park 904}
  \city{Amsterdam}
  \state{The Netherlands}
}
\email{e.kanoulas@uva.nl}

% By default, the full list of authors will be used in the page headers. Often, this list is too long, and will overlap
% other information printed in the page headers. This command allows the author to define a more concise list
% of authors' names for this purpose.

%
% The abstract is a short summary of the work to be presented in the article.
\begin{abstract}
%The increased capabilities of machine learning methods in recent years have stimulated real-life applications in which consequential decisions are made based on model predictions. Typically, t
The decision-making process of many state-of-the-art machine learning models is inherently inscrutable to the extent that it is impossible for a human to interpret the model directly: they are black box models. This has led to a call for research on explaining black box models, for which there are two main approaches. Global explanations that aim to explain a model's decision making process in general, and local explanations that aim to explain a single prediction. Since it remains challenging to establish fidelity to black box models in globally interpretable approximations, much attention is put on local explanations. However, whether local explanations are able to reliably represent the black box model and
provide useful insights remains an open question. We present Global Aggregations of Local Explanations (GALE) with the objective to provide insights in a model's global decision making process. Overall, our results reveal that the choice of aggregation matters. We find that the global importance introduced by Local Interpretable Model-agnostic Explanations (LIME) does not reliably represent the model's global behavior. Our proposed aggregations are better able to represent how features affect the model's predictions, and to provide global insights by identifying distinguishing features. 

%To what extent do local explanations reliably represent a models' behavior and how can the explanations best be used to gain global insights on a black box model? To answer these questions, 
\end{abstract}

%
% Keywords. The author(s) should pick words that accurately describe the work being
% presented. Separate the keywords with commas.

%
% This command processes the author and affiliation and title information and builds
% the first part of the formatted document.
\maketitle

\section{Introduction}
Over the last decade, many machine learning breakthroughs occurred, spiking widespread interest in the development of advanced machine learning methods, most specifically in the field of deep learning \cite{krizhevsky2012imagenet, srivastava2014dropout, greff2017lstm, goodfellow2014generative}. By using many layers of non-linear operations and abstractions, these complex models make it possible to make more accurate predictions than simpler methods can achieve. The increased capabilities of these machine learning models have stimulated the development of real-life applications in which consequential decisions are made based on model predictions. The downside of advanced machine learning models is that the decision-making process of such models is inherently inscrutable to the extent that it is impossible for a human to interpret the model directly: they are black box models. Since there is always a divergence between optimization goals and requirements in real-life applications, we cannot assume that the right rule is applied by the model, that the model takes in all - and only - the relevant information, and that all the data is accurate. The inscrutability of black box models, combined with the potential applications for consequential decisions that might affect our safety, our economy, or our opportunities, elicited a call for research on explaining black box models. Recently, this also led to EU regulation on the right to ``meaningful information'' about the logic involved in automated decision-making\footnote{EU General Data Protection Regulation: \url{https://gdpr-info.eu/}}. Although there is discussion on the legal implication of the terminology, admittedly it implies a right to explanation \cite{selbst2017meaningful}. The difficulty lies in questions regarding what makes for a valid, reliable and useful explanation; what are the desiderata? 

Prior work on explaining black box models has focused on either global or local explanations, where global explanations aim to explain a model's decision making process in general, while local explanations aim to explain a single prediction specifically \cite{ribeiro2016should,guidotti2018survey}. Global explanations suffer from the trade-off between interpretability of the explanation model and fidelity to the black box model, i.e. the more comprehensible a simplified explanation is, the less faithful it can be to the complexity of the black box model. Local explanations solve this by being restricted to local fidelity: fidelity to the black box model in the vicinity of the instance examined. The drawback is that it is unclear in what way the inspected instance is representative of the global behavior of the model. Moreover, due to the locality of these explanations and its undefined coverage, there would be no way of saying something valid about even a quite similar instance \cite{ribeiro2016should,murdoch2019interpretable}. 

Our work intends to fill this gap by presenting Global Aggregations of Local Explanations (GALE), to understand to what extent local explanations are able to provide global insights on a black box model. For this purpose, we analyze explanations obtained through Local Interpretable Model-agnostic Explanations (LIME) on models trained for a binary sentiment analysis task and a multiclass document classification task. We present several approaches to aggregate a set of local explanations and assess to what extent they are representative of the models' global decision rules and provide reliable and useful insights.

%Although local explanations are reliable by restricting their explaining power to local model behavior, a shortcoming is that it is unclear how this relates to global model behavior. 

\section{Related Work}
We first discuss various local explanations approaches that are within the scope of our research. Next, we elaborate on the the gap between local explanations and global model behavior,  and discuss previous work that addresses this limitation.

\subsection{Local Explanations}
The work of \cite{lundberg2017unified} identifies a family of approaches that provide a local explanation model in the form of a linear function of binary variables. By representing the features as binary variables, the weights $w$ of the linear model can be directly interpreted as feature attributions. This ensures that the explanation model is interpretable to humans, even though the original model might use complex features as input. For this reason, such methods require a mapping between $x$, the feature input of the model, and $x'$, an interpretable representation of that input. The explanation model $g(x')$ is a linear function of the interpretable representation, and approximates a model's prediction on an instance $f(x)$ as:

\begin{equation}
    g \left( x' \right) = w_0 + \sum_{j=1}^D w_j x'_j
    \label{eq:additive_feature_attribution}
\end{equation}

\noindent where $D$ is the input dimension of the instance explained. Local feature attribution methods provide the user with a local explanation through the weights of this linear model. The weights are considered feature attributions that reflect the influence on a prediction per feature. Attributions can either be supporting or opposing, and the higher the attribution, the higher it accounts for a feature's influence on the prediction. The sum of the attributions for all features in the input approximates the model output $f(x)$. 

Lundberg and Lee \cite{lundberg2017unified} propose a framework that unifies a family of approaches that provide an explanation model that is a linear function of binary variables as shown in Equation \ref{eq:additive_feature_attribution}. Methods that adhere to Equation \ref{eq:additive_feature_attribution} include Local Interpretable Model-agnostic Explanations (LIME) \cite{ribeiro2016should}, Layer-wise Relevance Propagation (LRP) \cite{bach2015pixel}, DeepLift \cite{shrikumar2017learning} and SHapley Additive exPlanation Values (SHAP) \cite{lundberg2017unified}.

LIME offers a local explanation by approximating the model in the vicinity of the instance being explained. It gathers local information by sampling instances from the instance being explained and is therefore considered a perturbation-based approach. The linear model that is fit on these samples offers an interpretable explanation for a specific prediction that adheres to Equation \ref{eq:additive_feature_attribution} \cite{ribeiro2016should, lundberg2017unified}. 
LIME is a model-agnostic approach to the extent that it can explain any classifier. Beyond this, the approach has inspired research on perturbation-based explanations for other types of tasks, such as models for ranking \cite{ter2018faithfully}.

Layer-wise Relevance Propagation is a backpropagation-based approach. It computes a relevance score, i.e. attribution, for each input neuron by redistributing the output of the prediction function backwards layer-by-layer. The exact redistribution function can differ depending on the classifier, as long as it satisfies the \textit{relevance conservation} property: at each layer the total amount of relevance, i.e. the sum of relevance of all neurons in the layer, equals the output of the prediction function \cite{bach2015pixel}.

DeepLift is another backpropagation-based approach that improves over LRP by introducing the notion of a \textit{reference value}, which is defined as the neurons activation to a reference input. The reference input is defined per task for the input neurons and propagated through the network to obtain reference values for all neurons. DeepLift obtains contribution scores, i.e. attributions, by expressing the difference from reference value of the output neuron in terms of the difference from reference value of the input neuron \cite{shrikumar2017learning, lundberg2016unexpected}.

\citet{lundberg2017unified} propose to use the concept of \textit{Shapley values}, from the field of game theory, to quantify feature importance. The Shapley value of a feature is the averaged marginal contribution of that feature to all possible subsets of features, i.e. meaning the average difference in prediction with or without the feature included for each subset. \citet{lundberg2017unified} present several approaches for approximating the Shapley values to obtain an explanation model, calling their approach SHapley Additive exPlanation values (SHAP).

The evaluation of our approach that is presented in this paper uses explanations obtained by LIME. Nonetheless, the global aggregations that we present in Section \ref{section:global_aggregations} are applicable to all local explanations that adhere to Equation \ref{eq:additive_feature_attribution}. Throughout this paper we will use the more general description \textit{local explanation} to refer to this specific family of approaches.

\citet{ribeiro2016should} stipulate LIME suffers from the fact that it is unclear what the coverage of an explanation is, i.e. to what extent it generalizes to other situations. Correspondingly, \citet{mittelstadt2018explaining} claim a three way trade-off, adding the size of the domain described as a third desiderata besides fidelity and interpretability. In another recent review, it was argued that explanations should be required to show relevance, meaning they must provide insights for users into a chosen domain problem \cite{murdoch2019interpretable}. In general, \citet{murdoch2019interpretable} state this requires a diversity in approaches, varying the balance in fidelity and interpretability, to meet the need of distinct domain problems. Particularly in the case of local explanations, they point out the gap between local explanations and global model behavior, leading to a call for future work to answer the question: to what extent local methods capture a model's behavior and how the explanations can best be used? Our work intends to increase the understanding of the limitations of local explanations and in doing so aims to improve the usefulness of such explanations.

\subsection{Global Insight From Local Explanations}

The authors of the LIME method acknowledge the gap between their method and the model's global behavior \cite{ribeiro2016should}. It is with this limitation in mind that they propose their \textit{submodular pick} algorithm. In order to provide global insight, this algorithm selects a set of data instances for which to present explanations to a user. It selects the instances by picking those that contain features with a high global importance score. 

\citet{ribeiro2016should} aim to define global importance such that features that explain many different instances have a higher global importance score, than features that explain less instances. In the case of text classification, they propose the global importance $I_j$ of a feature $j$ as the square root of the sum of its attributions, as shown and further elaborated on in section \ref{section:lime_importance}. Additionally, they aim to pick a subset $S$ out of $N$ instances, such that there is little redundancy in the features shown, arguing that if a feature appears in multiple instances, this would lead to similar explanations shown to a user. Both these intuitions are formalized as a coverage function $c$. Let $W$ be the $N \times M$ matrix containing the attributions per instance for each feature out of $M$ unique features. Then, given a set of instances $S$, the explanation matrix $W$ and the importance vector $I$, the coverage of the set $S$ is defined by: 

\begin{equation}
    c \left( S, W, I \right) = \sum_{j=1}^{} \; \lbrack \exists i \in S : W_{ij} > 0 \rbrack \; I_j
\end{equation}

\noindent Maximizing this function would yield the optimal set of instances covering important features. However, since this problem is NP-hard, a greedy algorithm is proposed that iteratively adds instances to the set $S$ to approximate the optimal set. The instance $i$ added at each iteration is the one with the highest marginal gain, which is defined as $c \left( S \cup \{i\}, W, I \right) - c \left( S, W, I \right)$ \cite{ribeiro2016should}. 

In the paper proposing this approach \cite{ribeiro2016should}, only limited evaluation on the submodular pick is provided. The authors only compare their approach to providing randomly selected instances to the user. They show that providing explanations for specific instances chosen by the submodular pick algorithm makes users more able to make choices that positively affect performance, such as comparing between models and features, than when they are shown a random set of instances. It is unclear to what extent the chosen instances are representative for the global behavior of the model, and the importance function is not further evaluated. Therefore, the gap between the local explanations and the global model behavior remains.

%%%%%%%%%%%%%%%%%%%%%%%%%%%%%%%%%%%%%%%%%%%%%%%%%%%%%%%%%%%%
%                       METHODOLOGY                        %
%%%%%%%%%%%%%%%%%%%%%%%%%%%%%%%%%%%%%%%%%%%%%%%%%%%%%%%%%%%%
\section{Methodology}
\label{section:global_aggregations}

Our work intends to fill the gap between local explanations and global model behavior. For this purpose, we propose a set of Global Aggregations of Local Explanations (GALE). Rather than validating local explanations independently, we assess the emerging insight from aggregating multiple local explanations. The way to aggregate local explanations is not straightforward since the attribution scores are not determined in relation to other data instances. It is unclear how attributions between different instances or in support of different classifications relate to each other. This is further complicated when applied to a textual task, given the sparsity of features in this domain. Any choice of aggregation function implies assumptions about the way in which local explanations are representative of the global model behavior. In our discussion of GALE we intend to make these assumptions explicit. 

%This gap limits the reliability and usefulness of local explanations, because it is unclear to what extend these explanations capture a models' behavior and how they can be interpreted.

\subsection{Global LIME Importance}
\label{section:lime_importance}
Ribeiro et al. \cite{ribeiro2016should} propose their submodular pick algorithm to select a set of instances to show a user in order to provide global insight. In order to select a representative and informative subset of instances, they propose a global aggregation function $I$ to assess global feature importance. Specifically for the text domain, they define the global feature importance $I_j^{L\textsc{ime}}$ as the square root of the sum of attributions $W_{ij}$ of the feature $j$ over all data instances $i \in N$:

\begin{equation}
    I_j^{L\textsc{ime}} = \sqrt{ \sum_{i=1}^{N} | W_{ij} |}
\end{equation}

Two assumptions underlie this aggregation function: 
\begin{itemize}
    %\bfseries
    \itshape
    \item [\textbf{A1}] Features with higher attributions are expected to have a larger effect on model predictions than features with lower attributions.
    \item [\textbf{A2}] Features that occur more often are expected to have a larger effect on model predictions than features that occur less often.
\end{itemize}

\noindent In the aggregation over separate instances, these assumptions will not always hold. Although \textbf{A1} seems reasonable amongst the features within one instance, this is less certain for feature attributions from various instances. Since the explanation model is a linear function of attribution values, the magnitude of attributions are affected by the amount of features per explanation. Similarly, the magnitude of attributions are affected by the prediction value that the explanation model is approximating. However, in comparison between attributions from different instances, the absolute value of the attribution might be less informative than its relative importance within the instance. 

With respect to \textbf{A2}, the amount of occurrences of a feature might be a misleading notion for several reasons, especially in text classification. Firstly, the assumption is that occurrences across different instances amount to a higher influence of the feature. Common words such as ``the'', ``and'', or ``is'', are thus likely to be ranked as very important due to many occurrences in different instances, even when their attributions are low. Secondly, as mentioned in \cite{ribeiro2016model}, local explanations for different instances can also be inconsistent with each other. This issue is further amplified in case of a multiclass classification task, because a single explanation does not show the features' possible relationship to other classes when providing a single example. Features that occur in documents of different classes will have a high global LIME importance, independently of which classes the individual attributions support, or whether the occurrences have a large impact on predictions.  

%%%%%%%%%% NO EXAMPLE
%An example of an inconsistency is shown in Figure \ref{fig:3} where the feature ``awfully'' is explained as being supportive for predicting a positive sentiment in the instance shown in Figure \ref{fig:3a}. From the example in Figure \ref{fig:3b} we know that the feature also - and maybe to a greater extend - influences the model in favor of negative sentiment predictions. An explanation in the form of feature attributions does not show a features' influence in other cases when providing a single example. This issue is further amplified in case of a multiclass classification task, because a single explanation neither shows the features' possible relationship to other classes when providing a single example. Features that occur in documents of different classes will have a higher global LIME importance, independently of what classes the individual attributions support.

\subsection{Global Average Importance}
In what follows we will not relax \textbf{A1}, but we will focus on \textbf{A2}. Whether features that occur more often in the data are more important, clearly depends on the domain. In case of textual data we often deal with sparse features; common words will occur often, while most other words will only occur in few instances. For this reason, we expect the global LIME importance to be unreasonably biased towards common words. 

Therefore, the first alternative aggregation we propose is the average importance, which is computed as the sum of attributions $W_{ij}$ averaged over the feature's occurrences in the dataset:

\begin{equation}
    I_j^{A\textsc{vg}} = \frac{\sum_{i=1}^{N} | W_{ij} |}{\sum_{i:W_{ij} \neq 0} \mathds{1}}
\end{equation}

Although the global average importance addresses the second assumption that is made in global LIME importance, it also makes its own assumption:
\begin{itemize}
    %\bfseries
    \itshape
    \item [\textbf{A3}] Features are expected to have a similar effect in all of their occurrences.
\end{itemize}

\noindent To understand why this assumption might not hold, imagine the case of a feature being important in the predictions for some class and less important when appearing in documents unrelated to that class. The occurrence in other documents will strongly lower its average importance, even though the feature was highly important for another class. 

\subsection{Global homogeneity-weighted importance}

The global homogeneity-weighted importance is designed to address \textbf{A2} and \textbf{A3}. The idea is to determine the homogeneity of a feature's influence on the model in order to deal with multiple occurrences and potential inconsistencies between occurrences. To quantify the homogeneity per feature, the spread of attributions over different classes is determined by Shannon entropy. First, we define $p_j$ as the vector of normalized LIME importance per class:

\begin{equation}
    p_{cj} = \frac{\sqrt{ \sum_{i \in S_c} | W_{ij} | }}{ \sum_{c \in L} \sqrt{ \sum_{i \in S_c} | W_{ij} | }}
\end{equation}

\noindent where $S_c$ is the set of all instances $i$ classified as class $c$ and $L$ is the set of class labels. The normalized LIME importance $p_j$ represents the distribution of feature $j$'s importance over all classes $c \in L$. The Shannon entropy of this distribution is defined by:

\begin{equation}
    H_j = - \sum_{c \in L} p_{cj} \log \left(p_{cj} \right)
\end{equation}

This entropy score is used to assess the degree of homogeneity with which the feature attributions of a feature are distributed over multiple classes. Low entropy indicates most of the attributions point to one particular class, as opposed to the case of high entropy in which attributions point to many classes. However, since entropy will be equally low for all features that only occur a single time in the test set, the entropy score does not discriminate in these cases. Therefore we propose to derive from this a homogeneity-weighted importance $I_j^{H}$. For this purpose, the entropy score is normalized and subtracted from $1$ to obtain a weighting factor that is close to $1$ if the feature is homogeneous and close to $0$ when its attributions are spread over many different classes. The homogeneity weighted importance is the LIME importance of a feature weighted by this weighting factor:

\begin{equation}
    I^H_j = \left( 1 - \frac{H_j - H_{min}}{H_{max} - H_{min}} \right) I^{L\textsc{IME}}_j
\end{equation}

\noindent where $H_{min}$ and $H_{max}$ are the minimum and maximum entropy measured across all features.

%%%%%%%%%%%%%%%%%%%%%%%%%%%%%%%%%%%%%%%%%%%%%%%%%%%%%%%%%%%%
%                   EXPERIMENTAL DESIGN                    %
%%%%%%%%%%%%%%%%%%%%%%%%%%%%%%%%%%%%%%%%%%%%%%%%%%%%%%%%%%%%
\section{Experimental Design}
\label{section:experimental_design}

Experiments are carried out on two distinct datasets: a relatively small sentiment analysis dataset and a larger document classification dataset. This makes us able to evaluate to what extent our approach is influenced by the complexity of the task and the amount of data available. For each task, a model with a task-specific architecture is trained on a subset of the data. Parameter tuning of these models is done based on a validation set that is not further used in our experiments. Subsequently, LIME explanations are obtained for the model predictions on a separate test set and used to gather the global aggregations over all instances in the test set. These are evaluated quantitatively based on our AOPC$_{global}$ metric and qualitatively by visualization of the important features according to each of the aggregations. In our experiments we answer the following research questions:
\begin{itemize}
    \itshape
    \item [\textbf{RQ1}] To what extent can global LIME importance represent how features affect the model's predictions?
    \item [\textbf{RQ2}] To what extent can global average and homogeneity-weighted importance represent how features affect the model's predictions, i.e. can the proposed aggregations improve on global LIME importance?
    \item [\textbf{RQ3}] To what extent do the quantitative results differ between a binary and a multiclass text classification task?
\end{itemize}

%%%% INCLUDE MODEL SOURCES?
% learning rate
% ADAM params momentum epsilon
\subsection{Sentiment analysis task}
Sentiment analysis is a binary classification task in which documents are labeled as expressing either an overall positive or negative sentiment \cite{pang2008opinion}. The sentiment analysis dataset used for this study consists of 3000 sentences with labels evenly distributed over a positive or negative sentiment\footnote{ Retrieved from \url{https://archive.ics.uci.edu/ml/datasets/Sentiment+Labelled+Sentences}}. \citet{kotzias2015group} selected these instances from larger datasets originating from the websites of IMDB, Amazon and Yelp, incorporating 1000 sentences per source. The LSTM architecture used for this task consist of one LSTM layer with tanh activation function and both input and recurrent dropout at $0.2$, followed by one fully connected softmax layer. The neural network is optimized with Adam over a run of $10$ epochs with a batch size of $32$. The input features are pretrained $100$-dimensional GloVe word embeddings\footnote{ Retrieved from \url{https://nlp.stanford.edu/projects/glove/}\label{foot:glove}}, which are not further fine-tuned during training \cite{pennington2018glove}. With this setup we obtain an accuracy of $0.85$.

\subsection{Document classification task}
The 20 Newsgroups dataset is a collection of approximately 20,000 newsgroup post, almost evenly distributed over 20 different classes\footnote{ Retrieved from \url{http://qwone.com/~jason/20Newsgroups/}}. The CNN architecture consists of three convolutional layers with ReLU activation functions, max pooling after each convolutional layer and dropout at $0.2$. The neural network has a final fully connected softmax layer, is optimized with Adam and run for $10$ epochs with batch size at $32$. The input features are pretrained $100$-dimensional GloVe word embeddings\footnote{ Retrieved from \url{https://nlp.stanford.edu/projects/glove/}}, which are not further fine-tuned during training \cite{pennington2018glove}. With this setup we obtain an accuracy of around $0.75$.

\subsection{Quantitative Evaluation}
\label{section:quantitative_evaluation}

In order to determine which global aggregation best represents the global model behavior, we propose an adaptation of the Area Over the Perturbation Curve (AOPC) evaluation. AOPC was proposed by \citet{samek2017evaluating} as an evaluation metric for local feature attribution methods. It defines a good local explanation as one that is able to identify the features that have the largest effect on model prediction. Adapting this, we propose AOPC$_{global}$ to evaluate to what extent the aggregations are able to identify the features that have the largest global effect on model predictions. The models' decisions on the resulting documents are evaluated by computing AOPC$_{global}$ over a range of consecutive feature removals and compared against a random baseline.

Our adaptation is that AOPC$_{global}$ is measured by progressively removing features per document in the order of their global ranking by GALE, as opposed to removing features in the order of their local ranking according to the local explanation as in AOPC for local evaluation. For each aggregation of features, the features in a document $x_{i}$ are ranked according to the global aggregation. Subsequently, features are iteratively removed from the original data point $x_{i}$ in the order of that ranking. Let $r$ be the vector of feature indices in document $x_i$ ranked in the order of the global aggregation. Then, the instance $x^{k}_{i}$ is the result of recursively removing the $k$ highest ranking features defined as follows: 
\begin{equation}
\begin{split}
x^{0}_{i} &= x_{i} \\
x^{k}_{i} &= s\left( x^{k-1}_{i}, ~r_{k} \right)
\end{split}
\end{equation}
where the function $s$ removes the $k^{\text{th}}$ highest ranking feature in document $x_i$ according to the global aggregation from index $r_k$ in the data point $x_i$. 

We propose AOPC$_{global}$ as a metric to evaluate the global ranking of features per aggregation by quantifying the effect of $K$ removed features on model predictions. It is defined as the the averaged cumulative sum up to $K$ of the drop in predicted class probability averaged over all instances in the test set:
\begin{equation}
AOPC_{global} = \frac{1}{K + 1} \bigg \langle \sum_{k=0}^K f(x^0_i) - f(x^k_i) \bigg \rangle^{i \in N}_{avg}
\end{equation}
\noindent where $\langle \cdot \rangle^{i \in N}_{avg}$ denotes the averaging over all instances in the test set and the black box prediction function $f$ returns the probability for the predicted class. 

This metric is computed over a consecutive range of removals per document $K$. Specifically, at $K=1$, AOPC$_{global}$ is evaluated by removing from each document one feature, that is the highest ranking feature in that document according to the global aggregation. The AOPC$_{global}$ at $K=2$ is computed by removing the two highest ranking features according to the global aggregation from each document, and so on. The resulting curve is evaluated on two aspects. Firstly, the overall height of the curve is assessed. A higher curve indicates a better ranking of features in the order of global influence on predictions. Secondly, we assess the initial slope of the curve. The steeper the slope of the curve for the first features removed, the stronger their influence on the model. A good aggregation is expected to demonstrate a positive decreasing slope for the AOPC$_{global}$ curve. 

\subsection{Qualitative Evaluation}
\label{section:qualitative_evaluation}

In addition to the quantitative evaluation, we present qualitative visualizations that enable examination of the global insights provided by GALE by demonstrating the most influential features per class according to each of the aggregations. In this perspective, a good global aggregation is one that considers features important if they distinguish between classes. Based on the qualitative visualizations we intend to answer \textbf{RQ1} and \textbf{RQ2} by observing whether the aggregations are able to identify distinguishing features. 

A slightly adapted version of each aggregation function is used to compute per class global aggregations. Let $S_c$ be the set of all instances $i \in N$ classified as class $c$. Then, the global LIME class importance for feature $j$ and class $c$ is defined as:

\begin{equation}
    I_{cj}^{L\textsc{ime}} = \sqrt{ \sum_{i \in S_c} | W_{ij} | }
\end{equation}

\noindent The global average class importance for feature $j$ and class $c$ is defined as:

\begin{equation}
    I_{cj}^{A\textsc{vg}} = \frac{\sum_{i \in S_c} | W_{ij} |}{\sum_{i \in S_c:W_{ij} \neq 0} \mathds{1}}
\end{equation}

\noindent And lastly, the global homogeneity-weighted class importance for feature $j$ and class $c$ is defined as:

\begin{equation}
    I_{cj}^H = \left( 1 - \frac{H_j - H_{min}}{H_{max} - H_{min}} \right) I^{LIME}_{cj}
\end{equation}

\noindent Notice that for the homogeneity-weighted class importance, the weighting factor remains the same as in Equation 7; it is still computed over all classes.

These global class importance functions are used to visualize the most important features per class, as determined by each of the aggregation methods. The selected features are plotted using t-SNE for dimensionality reduction on the word embeddings \cite{maaten2008visualizing}. More than just illustrating the most influential features, these visualizations present clusters of words that share a similarity in being indicative of a particular class. A global aggregation that can better identify distinguishing features than other aggregations, is expected to demonstrate more distinct clusters of words.
%The amount of features visualized is chosen per task and aggregation, since the maximum amount for which the visualization remains comprehensible could differ per case. 

%%%%%%%%%%%%%%%%%%%%%%%%%%%%%%%%%%%%%%%%%%%%%%%%%%%%%%%%%%%%
%                         RESULTS                          %
%%%%%%%%%%%%%%%%%%%%%%%%%%%%%%%%%%%%%%%%%%%%%%%%%%%%%%%%%%%%
\section{Results}

\subsection{Quantitative Results}
\label{section:quantitative_results}
%\textcolor{red}{maybe in this section you can highlight which RQs are answered. By adding something like, `this answers RQ1'.}

\begin{figure}[h]
  \centering
  \includegraphics[width=\linewidth]{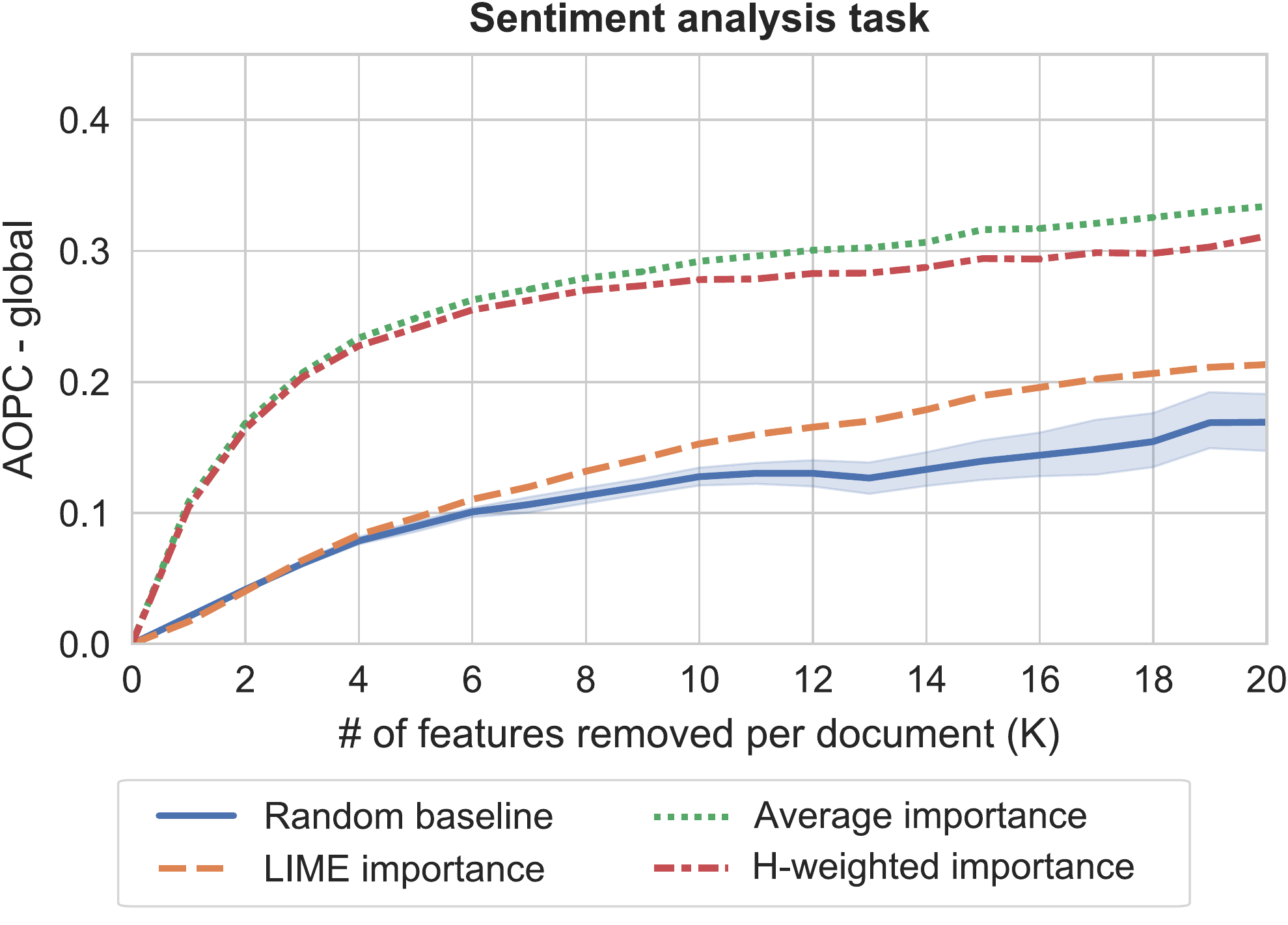}
  \caption{Quantitative evaluation of GALE on the sentiment analysis task. The AOPC$_{global}$ @ K is presented over a range of feature removals $K$ up to $20$.}
  \Description{AOPC$_{global}$ @ K on sentiment analysis task.}
  \label{fig:sa_acpd}
\end{figure}

\begin{figure}[h]
  \centering
  \includegraphics[width=\linewidth]{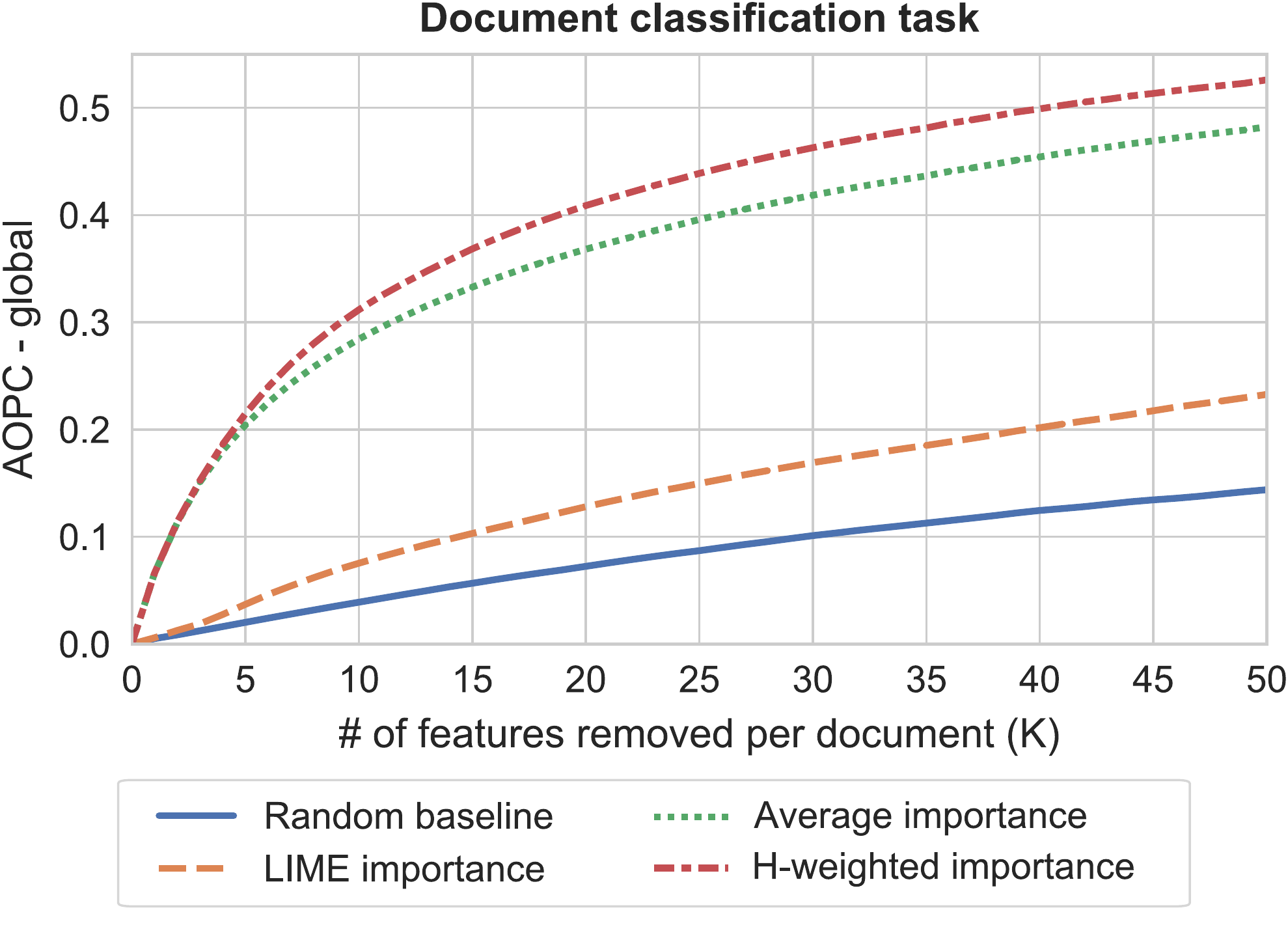}
  \caption{Quantitative evaluation of GALE on the multiclass document classification task. The AOPC$_{global}$ @ K is presented over a range of feature removals $K$ up to $50$.}
  \Description{AOPC$_{global}$ @ K on multiclass document classification task.}
  \label{fig:ng_acpd}
\end{figure}

We evaluate if features that are considered globally important according to the aggregations, indeed have a large impact on model predictions. For each of the global aggregations, features are removed in order of global importance, and compared against a random baseline for which the removed words are selected at random. The random baseline is averaged over five runs; the variance is also shown in the result plot. The evaluation is carried out for both classification tasks described in Section \ref{section:experimental_design}. The results for the sentiment analysis task are shown in Figure \ref{fig:sa_acpd}, where up to $20$ features are removed per sentence. Figure \ref{fig:ng_acpd} presents the results for the 20 Newsgroup text classification task. Since this task entails larger documents, results are shown for $K$ up to $50$ features removed per document. 

In both Figures \ref{fig:sa_acpd} and \ref{fig:ng_acpd}, it can be seen that the global LIME aggregation obtains only slightly higher AOPC$_{global}$ results compared to the random baseline. In particular it is found that the initial steepness of all LIME importance curves is equal to the initial steepness of the baseline. This indicates that the global LIME aggregation especially fails to correctly identify the most important features; the first features removed in the evaluation affect the models' predictions and performance no more than average. Our findings indicate that global insights through aggregation can be improved by selecting an aggregation function that better represents the local explanations with respect to the global model behavior. 
% Global LIME importance performs only slightly better than the random baseline, indicating that the global LIME importance aggregation is not able to reliably represent how features affect the models' global decision making process.

Both the global average importance and the global homogeneity-weighted importance surpass the AOPC$_{global}$ values obtained by global LIME importance and the random baseline, demonstrating that these aggregations are more able to reliably represent the model's global behavior. Additionally, for both aggregations, the AOPC$_{global}$ displays a much steeper initial slope of the curve. This finding implies that the global average  and global homogeneity-weighted importance, more adequately identify the most important features; these aggregations are better able to rank the features based on their global influence on model predictions. This is evidence for our hypothesis that the global LIME importance is based on misleading assumptions. More general, it reveals that the choice of aggregation function matters regarding its ability to represent the model's global decision making process.
% Global average importance and global homogeneity-weighted importance perform better than global LIME importance, showing they are better able to rank features in the order of their global influence on the model. This indicates that these aggregations better represent the models' global decision making process.

The average importance aggregation performs slightly better on AOPC$_{global}$ than the homogeneity-weighted importance in case of the sentiment analysis task. On the contrary, the homogeneity-weighted importance displays a steeper curve for AOPC$_{global}$ in the document classification task. Taking into account the difference in scaling between these tasks, the difference in the document classification task is stronger than in the sentiment analysis task. A possible explanation for why homogeneity-weighted importance performance better on the document classification task is that the sentiment analysis task is a binary classification task, while the 20 Newsgroup classification is a multiclass classification task. In a binary classification a local explanation for a particular instance informs about the feature influence for all possible class predictions - there are only $2$ classes. The attribution for a feature is either in support of the predicted class or against it, in the latter case this signifies support for the opposing class. In the case of multiclass classification, local explanations only provide an explanation for the influence of features in light of the predicted class. The global average importance of a feature that is influential for some classes would be significantly lowered due to low attributions in explanations for other classes. The global homogeneity weighting factor is more appropriate when explaining a multiclass classification model, because the weighting factor is affected by the spread of a feature's attributions over different classes, specifically to the degree of uniformity of that distribution. The effect is that global importance is reduced more for features that obtain high attributions for multiple classes than for features with high attributions for one class and low attributions in explanations for other classes.
% Global homogeneity-weighted importance best represents the models' global decision making process for the multiclass document classification task. There is only a small difference in performance between global average importance and global homogeneity-weighted importance in the binary sentiment analysis task.

\subsection{Qualitative Results}
\label{section:qualitative_results}

To deepen our understanding of which global aggregation best provide global insight in a complex model, several visualizations are presented as described in Section \ref{section:qualitative_evaluation}. In this paper we provide a confined version of the qualitative evaluation. More elaborate results are available at \url{https://github.com/iwcvanderlinden/GALE}. 
% MORE CAN BE FOUND ONLINE
\begin{figure}[hbtp]
  \centering
  {\includegraphics[width=\linewidth]{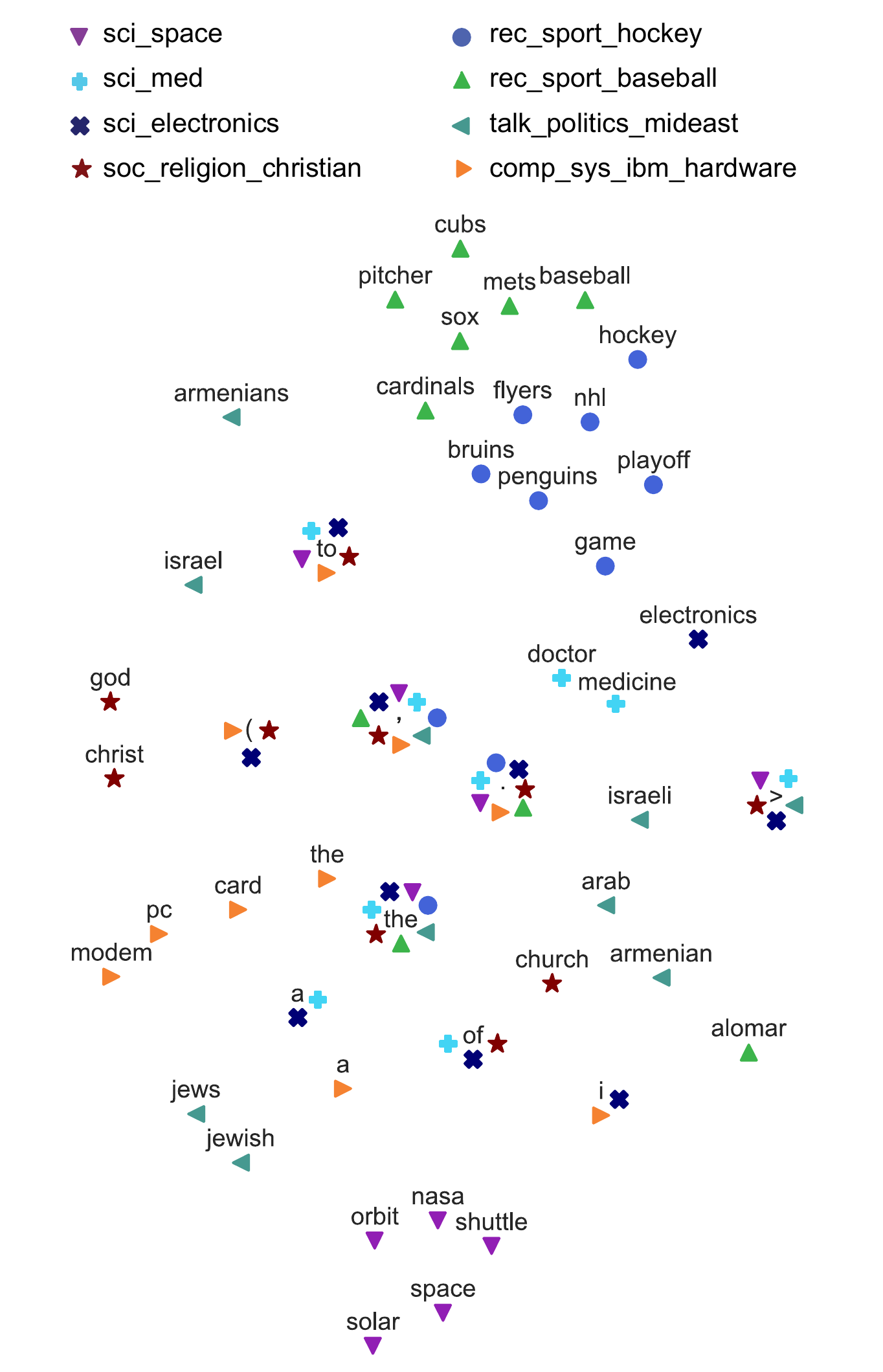}}
  \caption{Top 10 features per class according to global LIME importance. Class-specific clusters can be recognized for the hockey, baseball and space classes. On the other hand some clusters consist of the same common word or punctuation mark considered important for multiple classes.}
  \Description{Top 10 features per class according to global LIME importance.}
  \label{fig:lime}
\end{figure}

Figures 3-5 demonstrate the top ten features per class for 8 out of 20 Newsgroups classes, according to each of the global aggregations respectively. Firstly, the presented visualization for global LIME importance contains class-specific clusters of distinguishing features, as well as less substantive features, e.g. common words and punctuation, that do not appear in clearly distinct clusters. Both features that are likely and unlikely to distinguish between classes, are deemed important by the global LIME aggregation. Secondly, the visualization for global average importance contains class-specific clusters of distinguishing features for a minority of the classes in the document classification task. The global average aggregation considers substantive features important, i.e. no common words and punctuation. The qualitative visualization for global homogeneity-weighted importance demonstrates class-specific clusters of distinguishing features for a majority of the classes in the document classification task. Features deemed important by the global homogeneity-weighted aggregation are substantive features that distinguish between classes.

\section{Conclusion \& Future work}
In this work we propose GALE, which aims to provide insights in a black-box model's global decision making process. Overall, we conclude that Global Aggregations of Local Explanations have the potential to provide global insights from local explanations. In addition to this, our findings reveal that the choice of aggregation matters regarding the ability to gain reliable and useful global insights on a black box model. 

Our work offers opportunities to further develop GALE for different tasks as well as for other local explanations methods. For instance, regression and ranking tasks would require different aggregation functions. Future work could follow the procedure outlined in our methodology. Determine which assumptions are likely or unlikely to hold given the domain of the task, and design global aggregations accordingly. 

Furthermore, the GALE framework could be used to gather further understanding of local explanations. Information about the representativeness of individual explanations helps users comprehend and mitigate the gap between local explanations and global model behavior. We also intend on evaluating the application of GALE via a user-study.

\begin{figure}[hbtp]
  \centering
  \includegraphics[width=\linewidth]{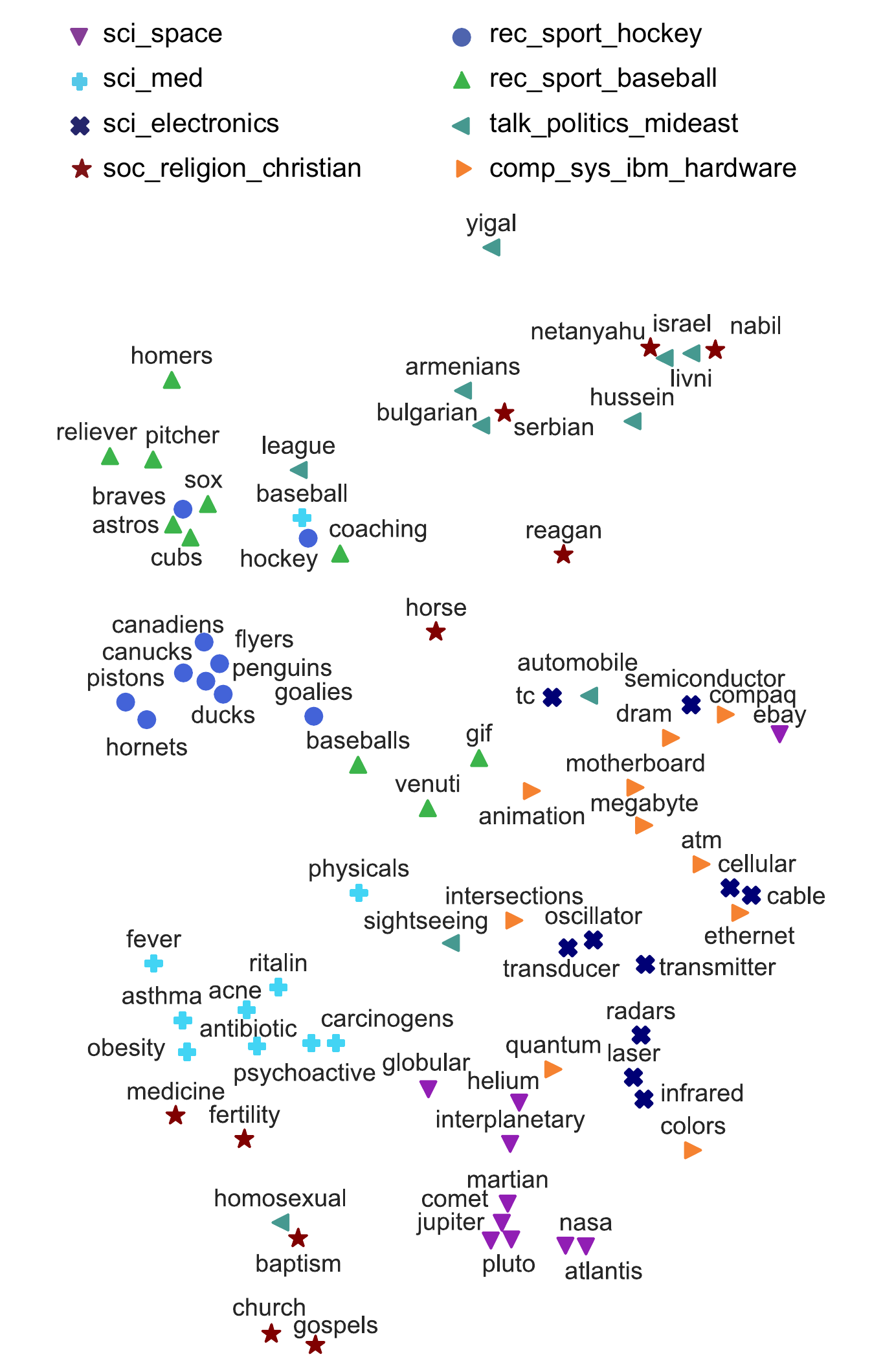}
  \caption{Top 10 features per class according to global average importance. Some class-specific clusters can be recognized for the hockey, baseball and space classes, most are not very distinct. The visualized features are more substantive than in Figure \ref{fig:lime}.}
  \Description{Top 10 features per class according to global average importance.}
  \label{fig:average}
\end{figure}

\begin{figure}[hbtp]
  \centering
  \includegraphics[width=\linewidth]{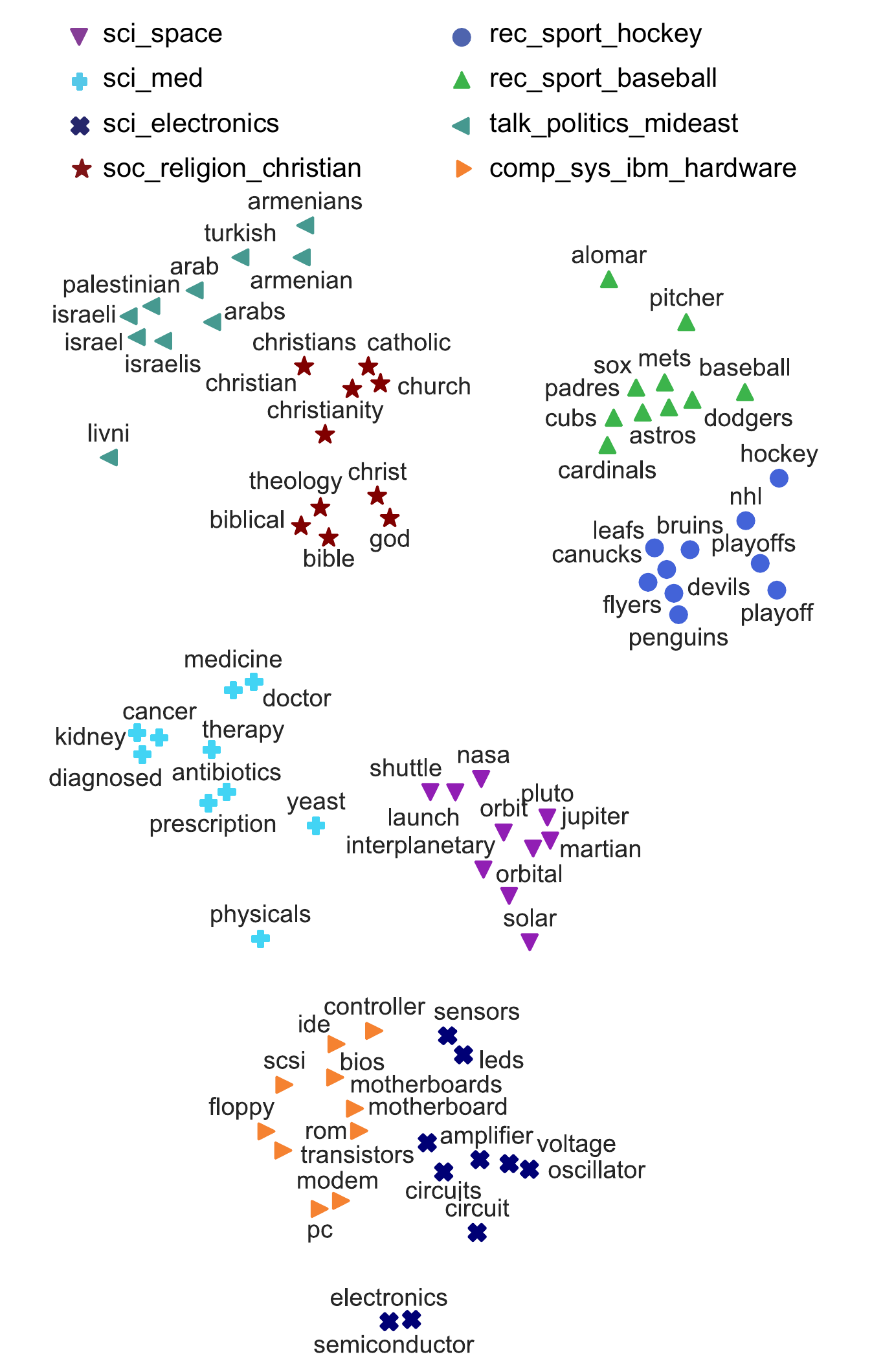}
  \caption{Top 10 features per class according to homogeneity-weighted importance. The visualized features are substantive and distinct class-specific clusters can be recognized for all classes.}
  \Description{Top 10 features per class according to homogeneity-weighted importance.}
  \label{fig:homogeneity}
\end{figure}

%\section{Acknowledgments}

%Identification of funding sources and other support, and thanks to individuals and groups that assisted in the research and the preparation of the work should be included in an acknowledgment section, which is placed just before the reference section in your document. 

%This section has a special environment:
%\begin{verbatim}
%  \begin{acks}
%  ...
%  \end{acks}
%\end{verbatim}
%so that the information contained therein can be more easily collected during the article metadata extraction phase, and to ensure consistency in the spelling of the section heading. 

%Authors should not prepare this section as a numbered or unnumbered {\verb|\section|}; please use the ``{\verb|acks|}'' environment.

%\section{Appendices}

%If your work needs an appendix, add it before the ``\verb|\end{document}|'' command at the conclusion of your source document. 

%Start the appendix with the ``\verb|appendix|'' command:
%\begin{verbatim}
%  \appendix
%\end{verbatim}
%and note that in the appendix, sections are lettered, not numbered. This document has two appendices, demonstrating the section and subsection identification method.

%
% The acknowledgments section is defined using the "acks" environment (and NOT an unnumbered section). This ensures
% the proper identification of the section in the article metadata, and the consistent spelling of the heading.

%
% The next two lines define the bibliography style to be used, and the bibliography file.
\bibliographystyle{ACM-Reference-Format}
\bibliography{references}

% 
% If your work has an appendix, this is the place to put it.
%\appendix

\end{document}